\documentstyle[aps,pre]{revtex}          
\tighten                                  
\begin{document}                          
\draft                                    
\twocolumn                               

\title{Dynamics of Spreading of Chainlike Molecules with
Asymmetric Surface Interactions}

\author{M. Haataja,$^{1,2}$ 
J. A. Nieminen,$^{2}$ and T. Ala--Nissila$^{1-3}$}

\address{
$^1$Research Institute for Theoretical Physics, 
\\P.O. Box 9 (Siltavuorenpenger 20 C), 
FIN--00014 University of Helsinki, Finland
}

\address{
$^2$Laboratory of Physics, Tampere University of Technology,\\
    P.O. Box 692, FIN--33101 Tampere, Finland}

\address{
$^3$Department of Physics, Brown University,
Providence, Rhode Island 02912, U.S.A.
}

\date{January 25, 1996}

\maketitle
\narrowtext

\begin{abstract}

In this work we study the 
spreading dynamics of tiny liquid droplets on
solid surfaces in the case where
the ends of the molecules feel different interactions 
with respect to the 
surface. We consider a simple model
of dimers and short chainlike molecules that cannot
form chemical bonds with the surface. We use
constant temperature
Molecular Dynamics techniques to
examine in detail the microscopic  
structure of the time dependent precursor film.
We find that in
some cases it can exhibit a high degree of local order
that can persist even for flexible chains.
Our model also reproduces the experimentally
observed early and
late--time spreading regimes where the
radius of the film grows $\propto t^{0.5}$.
The ratios of the associated transport coefficients
are in good overall agreement with experiments.
Our density profiles are also in good agreement with
measurements on the spreading of 
molecules on hydrophobic surfaces.

\end{abstract}



\section{Introduction}

Since the pioneering spreading experiments of
microscopic liquid droplets on surfaces  
by Heslot {\it {et al.}} \cite{Hes89a}, 
there has been increasing interest in phenomena 
occuring at microscopic lengthscales during
spreading. The experiments of 
Refs. \cite{Hes89a,Hes89b,Hes90,Alb92}
reveal that both the
molecular structure of the liquid and the type of
substrate used influence density
profiles of the droplets. For example, thickness profiles
of tetrakis (2--ethylexoxy)--silane and polydimethylsiloxane
(PDMS) droplets on a silicon wafer exhibit strikingly different
shapes under spreading \cite{Hes89a}. Tetrakis exhibits clearly
observable dynamical layering, while the spreading of PDMS
proceeds by a quickly evolving precursor layer of one molecular
thickness. 
Furthermore, the experiments of Valignat {\it {et al.}}
\cite{Val93} and Cazabat {\it {et al.}} \cite{Caz94}
address the important role of surface grafting, and
asymmetrical surface
interactions in particular, on the spreading dynamics.
For example, in Ref. \cite{Caz94} 
the density profiles for trisiloxane 
polyoxyethylene droplets spreading on a hydrophilic surface
resemble a ``sand pile'', whereas the same liquid forms a 
very compact bilayer on a hydrophobic surface.  

Despite drastic differences in the density profiles,
an interesting feature in the experiments of Refs. 
\cite{Hes89a,Hes89b,Hes90,Val93} is that the
time dependence of the 
radius of the precursor film $r(t)$ follows
``diffusive'' behavior $r(t) \sim t^{0.5}$ for
all times measured. Moreover,
the experiments of Valignat {\it et al.} \cite{Val93} 
report two distinct ``diffusive'' regimes
comprising a rapid early--time region 
followed by a considerably 
slower late--time one. Typical estimates
for the ratio between the corresponding
early--time and late--time
transport coefficients are of the order of 100. 
The emergence of the early--time regime can most 
simply be explained by assuming 
that the flux onto the surface is constant, {\it i.e.}
$dN_p(t)/dt=const.$, where $N_p(t)$ is the number
of molecules in the effectively
2D precursor film \cite{Haa95II}. From this it
follows that $r(t) \sim \sqrt{N_p(t)} \sim t^{0.5}$.
The late--time behavior is due to crossover
towards 2D diffusion 
that takes over in the submonolayer regime 
\cite{Alb92,Ala96}.

Motivated by these experimental discoveries 
a number of theoretical models have been proposed. 
However, progress has been rather moderate.
Analytic theories to date deal with dynamical
layering only \cite{Abr90,deG90}. Abraham {\it et al.} 
\cite{Abr90} considered a solid--on--solid 
(SOS) model of a layered 
droplet in rectangular geometry.
The width of the precursor film was found
to evolve linearily in time, implying that the effective
flux of the 
liquid into the precursor film is indeed constant in time.
De Gennes and Cazabat \cite{deG90} considered a model in which
the layers have already formed.
This model gives simple relations for the time--dependence 
of the radii of the layers. In particular, for a completely layered
droplet, the precursor film develops approximately ``diffusively''
in time \cite{diffusive}.
However, both models treat the liquid as structureless and
therefore cannot be applied to studying the effects of 
molecular structure or details of interactions
on the spreading dynamics.

For both coarse--grained and more microscopic models,  
a number of computer simulations have been performed 
\cite{Hei91,Yan91,Nie92,DeC93,Nie94,Ven94,Wag95,DeC95}
to study the dynamics of spreading. 
In particular, using Molecular Dynamics (MD) simulations
it was concluded in Ref. \cite{Nie94} that both 
the chain--like nature of the molecules and the chain--surface
interactions can significantly 
influence the structure of the precursor layer. Using a cylindrical
droplet geometry, Refs. \cite{Nie92,Nie94} 
reported an ``almost linear''
early time regime for the width $w(t)$ of their precursor
film (again indicating a constant flux), 
followed by a late--time diffusive region. Most recently,
the $t^{0.5}$ behavior was observed in another MD simulation
\cite{DeC95}. Qualitatively, the two time regimes
have been seen in Monte Carlo simulations, too 
\cite{DeC93,Ven94}.

In the present work, our aim is to employ MD simulation
techniques to study a particularly interesting aspect
of droplet spreading, namely the case where the fluid
molecules can feel asymmetric interactions with respect
to the surface. We will concentrate mostly on the
microscopic structure of the precursor film, 
its time dependence and the quantitative evaluation 
of the associated transport coefficients.
Our work is motivated by recent experiments
on such systems \cite{Val93,Caz94}, as well as
the practical importance of the molecular structure 
of thin layers.
In particular, a common way of controlling 
the surface energetics of a solid 
substrate is to use grafted molecules that adsorb on it,
sometimes forming chemical bonds and 
brushed layers \cite{Mil91}. Such surfaces have important
applications in {\it e.g.} coating and lubrication. 
Another interesting class of systems comprises amphiphilic molecules
such as detergents where a strong asymmetry of interactions
causes layered structures to form \cite{Isr92}.
Spreading dynamics of such molecules is then of
particular interest in trying to understand how these
layers form, and how well ordered they are.

The outline of this paper is as follows. First, we
introduce the model in detail.
Then we present results for the spreading
dynamics of molecules consisting of
two, four, and eight units with
an asymmetry of interaction with respect to the
underlying surface. Our results reveal that in some
cases the emerging precursor layer may exhibit a high
degree of local order that persists even for the
longer chains. This is reflected in the corresponding
density profiles, some of which bear a close resemblance
to the experimental ones \cite{Caz94}.
We also examine the time dependence
of the radius of the precursor layer, and find the
two different ``diffusive'' regimes in agreement with
previous studies \cite{Nie92,Nie94} and 
experiments \cite{Val93}. We furthermore
evaluate the corresponding transport coefficients
and find that their ratio is in good qualitative
agreement with the experiments. 
Finally, we briefly discuss the influence of the choice
of thermostat in our model. A brief account for the
results for dimers has been previously given in Ref.
\cite{Haa95}. 

\section{Model}

\subsection{Interactions} 

Our MD 
model is analogous to the one introduced in Refs. 
\cite{Nie92,Nie94}.
The \( n\)--mer molecules consist of \(n\) Lennard--Jones (LJ) 
particles. Within the chain, the LJ interaction between
them is purely repulsive, {\it i.e.} of the form 
$V_{LJ}^{intra}(r) =4 \epsilon_f(\sigma_f/r)^{12}$ to
prevent spatial overlap. The potential parameters are
$\sigma_f = 2.3$ {\AA} for the
width and $\epsilon_f = 0.1703$ eV for the depth,
respectively. Additionally, the particles are 
interconnected by a very rigid but orientationally isotropic 
harmonic oscillator pair potential 
\( V_{c}= \frac{1}{2} k (r-r_{0})^{2} \), 
where \( k=10000 \epsilon_{f} / \sigma_{f}^{2} \) and $r_0 = 2^{1/6} 
\sigma_f$ so that the chains do not easily stretch. 
There is also an 
angle dependent potential \( V_{\theta}=\epsilon_{\theta} 
(\cos \theta + 1) \), for \(n > 2 \). For our studies we examine 
two cases, namely \( \epsilon_{\theta}=10 \epsilon_{f} \) (rather 
stiff chains) and \( \epsilon_{\theta}=0 \) (completely flexible 
chains). 
There is no torsion dependent potential within a chain since we
consider linear chains. 

Interchain interactions are modeled by the following pairwise 
LJ interaction between $n$ LJ monomers:
\begin{equation}
V_{LJ}(r)=4 \epsilon_f[(\frac{\sigma_f}{r})^{12}-
(\frac{\sigma_f}{r})^6]\,\,\mbox{.}
\end{equation}
The substrate on the other hand 
is modeled by a flat continuum LJ material; 
it is thought to be homogeneous and its unit 
volume \( \Omega\) = 1. The total substrate interaction is 
obtained by integrating the LJ potential over the half space 
\( z \leq 0\) with the result 
\begin{equation}
V_{i}(z) = -\frac{A}{z^{3}} + \frac{B}{z^{9}}\,\,\mbox{,}
\end{equation}
where
\begin{equation}
A = (2 \pi/3)\rho_{s} \epsilon_{i} \sigma_{i}^{6}\,\,
\mbox{and}\,\,B = (4 \pi/45)\rho_{s} \epsilon_{i} \sigma_{i}^{12}
\,\,\mbox{,}
\end{equation}
and where $\rho_s$ denotes the number density of particles
comprising the substrate.
Different chain--surface interaction parameters employed in 
this study are presented in Table I and shown in Fig. 1.

The asymmetrical nature of the chain--surface interactions 
comes about through the choice of different set of surface 
interaction parameters for the grafted end as
compared to the 
other monomer units along the chain. The ``grafted'' end 
interaction is set to 
$V_{\mbox{I}}$ 
in every case studied in this work, whereas for dimers we employ 
$V_{\mbox{II}}$ ({\it ordinary case}) and $V_{\mbox{III}}$ 
({\it shifted case}), 
for tetramers $V_{\mbox{IV}}$ ({\it ordinary case}) and 
$V_{\mbox{V}}$ ({\it shifted 
case}), and for octamers $V_{\mbox{IV}}$. 
These potentials have been constructed in such a way
that in 
the {\it ordinary} case individual chains tend to lie 
{\it parallel} to the surface while in the {\it shifted} 
cases they lie {\it perpendicular} to it. It should be noted
that the chains in 
our model are not allowed to form chemical bonds 
with the surface.

\subsection{Choice of physical units}

The physical units are determined by the Hamaker
constant $A_H$ of the substrate and by the number
density of substrate atoms.
In our units, we have fixed $\rho_s=1.0 \, \mbox{\AA}^{-3}$. 
The Hamaker constant $A_H$ of the substrate fixes
the effective bond length $b_l$. 
This can be seen as follows: the Hamaker constant 
between two materials is defined by
\begin{equation}
A_H = 4 \pi^2\,\, \epsilon_s\,\sigma_s^6\,
\rho_s\,\rho_f\,\,\mbox{,}
\end{equation}
where $\rho_f$ denotes the number density of molecules in the fluid 
\cite{Isr92}. We have fixed $A_H$ through the choice
$\epsilon_s=2.8 \times 10^{-20}\,\mbox{J}$ and $\sigma_s=1.25 
\times 10^{-9} \,\mbox{m}$ and requiring that it 
is realistic, {\it{i.e.}} $A_H \sim 10^{-18} \,\mbox{J}$ 
\cite{Isr92}. 
This fixes the number density of the fluid to be $\rho_f 
\sim 10^{24} \,\mbox{m}^{-3}$. 
On the other hand, $\rho_f \sim b_l^{-3}$ and 
hence the effective bond length $b_l \sim 10^{-8} \,\mbox{m}= 100 
\,\mbox{\AA}$. This justifies the use of a flat, 
continuum substrate in our model studies.

For an LJ system there is a typical time scale, which 
is fixed by the choice of $\epsilon_f$, $\sigma_f$, and the mass 
$m$. In our bare units, the mass of a monomer is 
$m_b=63.5$ amu, $\epsilon_f =0.1703 \,\mbox{eV}$ and $\sigma_f=2.3 
\,\mbox{\AA}$. The combination that yields 
a quantity with the dimension of 
time is $\tau_c=\sqrt{(m_b \sigma_f^2)
/ \epsilon_f} = \sqrt{m_b/ \epsilon_f}\, \sigma_f$ which in
our bare units is $\approx 5 \times 10^{-13} \,\mbox{s}$. 
In our simulation algorithm, we have chosen the time step
to be $0.01 \tau_c$.
To obtain physical units we use the bond length $b_l
\approx 100\,\mbox{\AA}$ which scales $\sigma_f$,
and set the physical mass
of our effective monomers to be a realistic value of
$m=10^5\,\mbox{amu}$. Using these values to scale $\tau_c$
gives the time step  
in physical units to be $t_{r.u.} = 7.7 \times 10^{-13} \,\mbox{s}$.

\subsection{Choice of thermostat}

The dynamics of the system with a
Nos\'e--Hoover (NH) thermostat is
described by the usual equations of motion \cite{Nie94}:
\begin{equation}
\frac{d \vec{r}_{i}}{dt} = \frac{\vec{p}_{i}}{m_{i}}\,\,\mbox{,}
\end{equation}
\begin{equation}
\frac{d \vec{p}_{i}}{dt} = -\nabla_{i} V - \eta \vec{p}_{i}\,\,\mbox{,}
\end{equation}
and
\begin{equation}
\frac{d\eta}{dt} = [ \sum_{i} \frac{p_{i}^{2}}{m_{i}} - N_f k T_s] 
/ N_f k T_s \tau^{2}\,\,\mbox{,}
\end{equation}
where \(N_f\) is the number of degrees freedom, \(kT_s\) the temperature
for the thermostat, \(\eta(t)\) a time--dependent friction 
coefficient, and $\tau = 2.0 \times 10^{-14}\,\mbox{s}$
is a relaxation time. If we were to study
the equilibrium properties of our model,
we note that this choice of $\tau$
would remove any uncanonical temperature fluctuations
due to a hidden Toda demon \cite{Hol95}.
The equations of motion are solved using modified 
velocity Verlet algorithm (see {\it e.g.} Ref. \cite{AlTi}).
The simulations are performed at temperature $kT=0.8 \epsilon_f$ 
which is well above the triple point of an LJ fluid 
\cite{Nie94}. At the end of this work, we will also briefly
discuss the influence of choosing a local thermostat
based on Langevin dynamics to our results.

\subsection{Construction of the initial configuration}

We use the cylindrical geometry of Refs. \cite{Nie92,Nie94}
and construct an initial ridge--shaped droplet with periodic 
boundary conditions along the direction of the ridge which
is denoted by $y$. The 
length of the cylinder is 
$\approx 10 \, b_{l}$ for dimers, $\approx 16 \, b_{l}$ 
for tetramers and $\approx 38 \ b_{l}$ for octamers. 
Spreading takes place in the $x$ direction 
which lies perpendicular to the ridge. 
This choice of geometry 
is for computational convenience, and translating our 
results to the true 3D case is straightforward.

Constructing a proper initial configuration is complicated
by long relaxation times of the chain--like molecules.
To overcome this, 
in the beginning the locations of the end--groups of the 
\(n\)--mers are
chosen randomly. Then the chains are formed in such a way, 
that they are bent
90 degrees at each joint, while their directions are random. 
The initially rather sparse system is compressed to find the 
minimum of the internal energy. Then the droplet is allowed 
to equilibrate in the following way: the temperature of the 
system is set to \(0.1 \, kT_{s}\), where \(kT_{s}=0.8 \,
\epsilon_f\) denotes the actual simulation temperature. 
Then the temperature is raised to \(1.5 \, kT_{s}\) 
by continuously adjusting the temperature of the thermostat. 
All this time the system is allowed to evolve without 
the surface interaction. After this the temperature is lowered 
to \(kT_{s}\) in the same way; this should result in a 
configuration that is closer to the actual equilibrium 
configuration than the initial with. We have qualitatively
confirmed this by estimating the corresponding free energy
differences \cite{AlTi}.
The time scales used in constructing the initial configuration 
are comparable to the actual spreading simulation. 
After 
the system has been equilibrated, the substrate potential 
is ``switched on'' and spreading can take place. 

\subsection{Quantities calculated}

One of the main advantages of the MD method is that the
spreading dynamics can be followed in real time, and
detailed data on the configurations are available. 
To this end, we
have calculated the following quantities:

\begin{enumerate}

\def\theenumi{\roman{enumi}}
\def\labelenumi{(\theenumi)}

\item {\bf The width of the precursor film $w(t)$} is a measure 
of the average 
horizontal extent of the droplet in the $x$ direction. 
In our geometry the 
spreading takes place in the $x$ direction only and hence 
simple geometrical arguments give $w(t) 
\approx N_p(t)/(\rho L)$, where $N_p(t)$ is the number of 
particles in the precursor film, $\rho$ denotes the average 
number of particles per unit area in it
and $L$ denotes the length of 
the droplet along the $y$ axis. 
To convert this to the spherically symmetric 3D case we
assume that the flux of particles to the
precursor film $\propto t^{\alpha}$ where $\alpha
\approx 1$.
This immediately implies, that 
$w(t) \approx A t^{\alpha}$ for a compact layer.
It follows then that for our geometry 
\begin{equation} \label{equ1}
N_p(t) \approx A \rho L t^{\alpha} \,\,\mbox{.}
\end{equation}
On the other hand, 
the radius of 
the precursor film of a fully spherical droplet
that spreads radially, is given by
\begin{equation} \label{equ2}  
r(t) \approx \sqrt{N_p(t)/(\rho \pi)} \,\,\mbox{.}
\end{equation} 
Combining Eqns. (\ref{equ1}) and (\ref{equ2}) we obtain
\begin{eqnarray}
r(t) \approx \sqrt{A \rho L t^{\alpha}/(\rho \pi)} \nonumber \\
=\sqrt{AL / \pi} \,\,t^{\alpha /2} \nonumber \\
\equiv \sqrt{D_{e}} \,\, t^{\alpha /2} \,\,\mbox{,}
\end{eqnarray} 
where the associated early--time ``diffusion''
(transport) coefficient is defined by $D_{e}=LA/ \pi$.

\item {\bf The density profile} is of particular interest because 
it can be directly measured in the experiments. In our 
model it is defined as the average number of particles 
in a box of size \( \Delta x \Delta y\) taken at any 
fixed \(y\).

\item {\bf The pair correlation function} characterizes 
the degree of order in the system. 
It is defined as the following
configuration average \cite{AlTi}:
\begin{equation} \label{gr}
g(r,t) = (N/V)^{-2} {\Bigl \langle} \sum_{i} \sum_{j \neq i}
\delta({\vec {r_{i}}}^{\ cm}) \delta({\vec {r_{j}}}^{\ cm}- 
\vec {r} \,) {\Bigr \rangle} \,\,\mbox{,}
\end{equation}
where $N$ denotes the number of molecules in the system 
of volume $V$ and  $\vec {r}_{i}^{\ cm}$ is the position of the 
centre--of--mass of the $i^{\rm th}$ molecule. 
In particular, we have calculated $g(r,t)$
within the effectively 2D precursor film, where $N \equiv N_p(t)$
and $V \equiv A_p(t)$, the latter denoting
the time--dependent area of the precursor film.
It should be noted that 
since the system is not translationally invariant
in the $x$ direction, $g(r)$ starts 
deviating from its asymptotic values for large $r$, for which
$g(r) {\rightarrow} 1 \,\,\mbox{when}\,\, r {\rightarrow} \infty $.

\item {\bf The distribution of orientations for neighbouring chains 
$d_o$} is defined as follows: to every chain \(i\) a vector 
\(\vec {R}_{i}\) is assigned as \(\vec{R}_{i} = \vec{r}_{i+nl} 
- \vec{r}_{i} \). Here \(\vec{r}_{i}\) denotes the position of 
the grafted end of the chain \(i\), and \(\vec{r}_{i+nl}\) 
denotes the position of the other end of the chain. The 
normalized distribution is defined as follows:
\begin{equation}
d_o(\theta,t)=\frac{N_o(\theta,t)}{\sum_{\theta=0^{\circ}}
^{180^{\circ}}N_o(\theta,t)}\,\,\mbox{,}
\end{equation}
where $N_o(\theta,t)$ denotes the number of pairs of 
chains oriented at an angle $\theta$ relative to each 
other at time $t$. Only those pairs of chains whose 
centres of masses lie within a sphere of radius $2.5\,
\sigma_f$ are considered. This quantity  
indicates the spatial orientational correlations of 
neighbouring chains. This method is accurate when we 
use rigid chains, but in the case of flexible ones 
it gives only an approximate 
picture of the degree of orientational correlations
between neighbouring chains.

\item {\bf Distribution of bond angles $d_b$} is a time--dependent 
quantity that keeps track of bond angles at given times. 
The bond angle is defined to be the angle between 
consecutive bonds in a molecule and it is calculated 
by taking the dot product of the associated bond vectors. 
Bond vector is defined as the vector between consecutive 
monomers in the chain. Distribution of bond angles $d_b$ 
is defined as the following normalized histogram:
\begin{equation}
d_b(\theta,t)=\frac{N_b(\theta,t)}{\sum_{\theta=0^{\circ}}
^{180^{\circ}}N_b(\theta,t)}\,\,\mbox{,}
\end{equation}
where $N_b(\theta,t)$ denotes the number of bond angles 
with angle $\theta$ at time $t$.
It is calculated for the 
initial configuration and for a 
configuration taken at late stages of spreading. It is 
used in analyzing whether completely flexible chains 
become effectively stiffer in the course of spreading or 
not. This tendency would be revealed by a distribution 
profile peaked sharply around $\theta \approx 0^{\circ}$. 
This quantity is calculated for tetramers and octamers only.

\end{enumerate}

\section{Results}

First we consider short, rigid molecules with 
two monomer units. These are 
called {\it dimers}, and we will present results for
two different 
sets of surface interaction parameters. Then we consider 
longer flexible chains, and we present 
results for chains with four monomer units ({\it tetramers}) 
and with eight units ({\it octamers}). 

\subsection{Dimers}

In this section we present complete results for short 
and rigid molecules (see Ref. \cite{Haa95} for a brief summary). 
We have studied two different cases, 
namely one in which the equilibrium orientation of an 
individual chain is parallel to the surface (the {\it 
ordinary} case) and the other in which the molecules 
prefer to lie perpendicular to the surface (the {\it 
shifted} case). The shifted case can be considered
to be an effective model for rigid molecules with
one end hydrophobic and the other hydrophilic
relative to the surface \cite{Caz94}.
The grafted end interacts with the surface
with potential $V_{\mbox{I}}$ while
in the ordinary case the other end has $V_{\mbox{II}}$ 
and in the shifted case $V_{\mbox{III}}$.

\vspace{0.5cm}

{\bf The ordinary case}

\vspace{0.5cm}

Figs. 2(a) and (b) show a sequence of snapshots from a typical 
evolution of the droplet for $N=1525$ ordinary dimers
during spreading as seen along 
the axis of the ridge. The holes in the initial configuration 
are due to density fluctuations. 
The initial configuration 
is characterised by a disordered and liquid--like 
structure. This is revealed by the pair correlation 
function and the distribution of orientations for neighboring 
dimers for this geometry.

After switching on the surface attraction it can be seen 
that the dimers in the middle of the droplet that have 
not yet come into contact with the surface are mainly 
oriented perpendicular to the surface with the grafted 
end pointing downwards due to stronger surface attraction. 
The molecules that are on the surface, on the other hand, 
behave quite differently. It can be seen that the precursor 
film is very disordered except very close to the edges 
of the droplet. This is due to the high density of 
dimers near the center of the droplet which 
effectively prevents them from attaining 
their equilibrium orientations relative to the 
surface. At the edges of 
the droplet the dimers have enough room to lie 
flat on the surface. The final configurational 
stage is a thinning monolayer of molecules lying 
flat on the surface and exhibiting diffusive motion.

To quantify these observations we have calculated 
the pair correlation function of the center of
mass of the molecules within the precursor film. 
This case shown in Fig. 3 reveals that 
there is only weak short--range order characteristic of 
liquids. 
In order to characterize the degree of 
long--range orientational correlations
we have calculated 
the distribution of orientations for neighbouring dimers
at a late stage of spreading. 
It reveals that 
the overall orientational correlation between 
well--separated dimers is weak. This again is 
consistent with conclusions made above about the 
structural order within the droplet. 

Fig. 4 shows the evolution of the density profile of the 
droplet taken at three different time steps. 
It can be seen that the profiles are fairly smooth and 
rounded; the peaks are due to dimers that have not 
yet come into contact with the surface. 
At later times a step develops at the edge of the 
film where dimers tend to lie flat on the surface. 

Fig. 5 shows the dependence of the horizontal 
width of the precursor film
\( w(t)\) on time for a typical simulation run for $N=1525$. 
Qualitatively, the data look similar to that of Refs.
\cite{Nie92,Nie94} with two regimes visible. 
We have analyzed the data
as follows. First we assume that 
\begin{equation}
 w(t)= \left\{   \begin{array}{ll}
                 A(t-t_{1})^{\alpha}+w_0 \mbox{,} & \mbox{for 
$ t_{1} < t < t_{2}$,} \\
                 B(t-t_{2})^{\beta}+w_{0}' \mbox{,} & \mbox{for 
$ t>t_{2} $.}
                 \end{array}
       \right.
    \label{eqtype2}
\end{equation}
in which $t_{1}$ and $t_2$ denote cross--over times, 
$w_0$ and $w_0'$ are constants, and
$\alpha$ is the exponent for the first and $\beta$ 
for the second regime. 
>From the data one expects to 
find two different power law regimes. 
A standard trick is to estimate the 
initial transient time $t_{1}$ and 
then plot $ \ln \,(w(t)-w_0)$ {\it vs.} $ \ln\,t$. 
When trying to 
obtain $\beta$ in a similar manner, the difficulty 
lies in the fact that the estimation of $t_{2}$ 
which denotes crossover towards final 2D diffusion,
is not as straightforward as for $t_{1}$.  
A linear least--squares fit for the first
regime 
gives $w(t) \sim t^{0.8 \pm 0.1}$. 
The same procedure is then applied to the other regime.
The slope of the least--squares linear 
fit for the second ``diffusive'' regime gives $w(t) \sim t^{
0.5 \pm 0.1}$. 

These two exponents obtained from the data are 
consistent with the results of Refs. 
\cite{Nie92,Nie94}. 
They found a crossover for the 
width of the precursor film $w(t)$ from ``almost 
linear'' ($\sim t^{0.9}$) to ``diffusive'' ($\sim 
t^{0.5}$) behavior. When we translate this to the 3D 
situation, we recover the two ``diffusive'' regions 
with different effective transport coefficients in 
accord with experiments \cite{Hes89a,Hes89b,Hes90,DeC93}. 
Using our result $\alpha \approx 0.8$ and extracting $A$
we find that the early--time diffusion 
coefficient $D_{e} \approx 5.4 \times 10^{-6}\,\mbox{m}^2/\mbox{s}$. 
For the late time ``diffusion'' coefficient
we find $D_{\ell} \approx 1.2 \times 10^{-6}\,\mbox{m}^2/\mbox{s}$
and thus $D_{e} / D_{\ell} \approx 5$.
These values are somewhat larger than the measured ones that 
range between $0.4 - 2.0 \times 10^{-12}\,\mbox{m}^2/\mbox{s}$
\cite{Val93}. Also, typical experimental ratios are of the
order of $100-1000$.
The difference is not
surprising since in our units $T \approx 
1600 \,\mbox{K}$, and there are no surface diffusion barriers.
Corrugation of the surface would most
likely tend to lower the late time ``diffusion''
coefficient thereby making the ratio larger.
We note that extrapolating our values of $D$ to room
temperatures gives about $10^{-15}$ m$^{2}$/s, in
reasonable agreement with experiments.

The fact that $w$ is a function of both time $t$ and 
the number of dimers $N$ suggests that there might 
exist a scaling form for $w(t,N)$ as suggested in 
Ref. \cite{Nie94}, which is of the following form:
\begin{equation}
w(t)=t^x \Phi(t/N^y)\,\,\mbox{.}
\end{equation}
However, for the present case we find no 
such scaling for the range of times and system 
sizes studied. This is in part because for our 
relatively small systems, crossover to diffusive 
behavior is very sharp and thus the data do not collapse.

\vspace{0.5cm}

{\bf Shifted case}

\vspace{0.5cm}

For the shifted case, Figs. 6(a) and (b) show a typical 
evolution of the droplet for $N=1525$ shifted dimers
during spreading. Initial 
configuration is identical to that of the ordinary case.
Again, the dimers that have not yet come into contact 
with the surface behave in a manner similar to the 
ordinary case. The molecules in the middle of the droplet 
are mainly oriented perpendicular to the surface. 
A striking difference between the ordinary and the 
shifted case is the development of a compact 
precursor layer which appears very well ordered at 
all stages of spreading. Fig. 7 shows the pair 
correlation function within the precursor film 
for the shifted case taken at $t=80000 \, \mbox{r.u.}$.
Clear peaks can be observed corresponding up to 
about fourth or fifth nearest neighbor dimers. 
The precursor layer in this case indicates a high 
degree of local ordering even at these elevated 
temperatures.

We have also calculated the 
distribution of orientations for neighbouring 
dimers for the shifted case. The overall shape 
of the profile, which is shown in Fig. 8 
corresponding to late stages of spreading, 
reflects the properties discussed above. A 
clear peak can be seen indicating orientation of 
nearby dimers in a preferred (vertical) direction. 
In this case the orientational correlation of the 
dimers extends through the entire precursor layer. 

Fig. 9(a) shows typical density profiles of 
dimers taken at three different times. Whereas 
in the ordinary case a step developed at the edges 
of the droplet, in the shifted case the edge of the 
precursor film always remains very sharp and 
well--defined.
It is interesting to compare the calculated
density profiles with the experimental ones shown
in Fig. 9(b) for
triloxane polyoxyethylene molecules
spreading on silica bearing a dense grafted layer
of trimethyls \cite{Caz94}. In this case
grafting results in a hydrophobic surface.  
The triloxane polyoxyethylene is a hammer--shaped
molecule which 
has a hydrophobic (the trisiloxane head) and
a hydrophilic (polyoxyethylene tail) part.
The attraction of the hydrophobic group to the surface
and the repulsion between the hydrophilic part and the
surface forces the molecules lie perpendicular to the surface.
Despite the enormous difference in the horizontal 
scales, the simulated and experimental
profiles are in good qualitative agreement.

Fig. 10 shows the dependence of $w(t)$ for the shifted 
case. Again a crossover form ``almost linear'' to 
``diffusive'' behavior can be seen. The crossover 
appears to be somewhat sharper in this case. With 
the smallest system size studied the spreading stops 
completely due to finite--size effects. With larger 
systems the spreading continues 
in a diffusive manner. We have analyzed the data for
$w(t)$ in the same way as before, and find that
within the ``almost linear'' regime 
$w(t) \sim t^{0.8
\pm 0.1}$ and in the ``diffusive'' regime 
$w(t) \sim t^{0.4 \pm 0.1}$. 
We again convert these results into the 3D case 
and recover the two diffusive 
regions. We estimate that $D_{e} 
\approx 5.4 \times 10^{-6}\,\mbox{m}^2/\mbox{s}$, and
$D_{\ell} \approx 1.1 \times 10^{-7}\,\mbox{m}^2/\mbox{s}$
which give $D_{e}/D_{\ell} \approx 50$,
in good agreement with experiments \cite{Val93}.

The reason for $D_{\ell}$ being about an order of magnitude 
smaller than in the ordinary case can be understood
from simple energetic arguments.  
The activation energy 
for diffusion in the shifted case is larger due 
to the high degree or local ordering. 
We have estimated the activation energies $E_a$ for 
diffusion for the two cases by calculating the average energy
due to neighbours of a dimer located within two bond 
lengths from the edges of the droplet. 
In the ordinary 
case we find $E_{a}\approx 0.9\,\mbox{eV}$ and in 
the shifted case $E_{a}' \approx 1.2\,\mbox{eV}$.
If we make the assumption that $D_{\ell}$ follows 
the Arrhenius form $D_{\ell} \propto 
e^{-E_a/kT}$, we can estimate that 
\begin{equation}
\frac{e^{-E_{a}/kT}}{e^{-E_{a}'/kT}} 
\approx \,10 \,\,\mbox{.}
\end{equation}
This is fully consistent with 
results extracted from  
the width of the precursor film. 

In the case of shifted dimers we again checked the
scaling form of Eq. (16), but did not find a good data
collapse for the present times and system sizes studied.

\subsection{Tetramers}

For chains consisting of four monomers
we have studied 
three different cases, namely two ordinary 
cases (rather stiff and completely flexible tetramers) 
and a completely flexible shifted case. 
The surface potential for the grafted end is set to 
be $V_{\mbox{I}}$ 
and for the ordinary cases we employ the surface 
potential $V_{\mbox{IV}}$. In the shifted case the surface 
potential of the other end is set to $V_{\mbox{V}}$, whereas 
the monomers in the middle of the chain do not have any 
interactions with the surface. 

\vspace{0.5cm}

{\bf Rather stiff tetramers}

\vspace{0.5cm}

Rather stiff tetramers are characterised by a fairly 
strong angle--dependent potential between consecutive 
bonds in a chain, namely  \( V_{\theta} = 
\epsilon_{\theta} (\cos \theta +1) \), where 
\(\epsilon_{\theta}=10 \,\epsilon_{f} \). 
Figs. 11(a) and (b) show the evolution of a droplet 
for $N=785$ in a typical 
simulation run. The chains appear to be initially fairly 
straight and the calculation of the distribution 
bond angles confirms 
this observation. The apparent holes in the droplet 
are due to thermal fluctuations. The calculation of 
distribution 
of orientations for neighbouring chains also supports 
the conclusion that 
the structure of the initial configuration is 
disordered and liquid--like.

Qualitatively, the results are similar to the case of
ordinary dimers. The chains 
in the middle of the droplet are oriented 
approximately perpendicular to it, the grafted end being 
closest to the surface. Closer to the edges, the chains
tend to attain a more horizontal orientation. 
The centre of the droplet appears to have a very 
complicated structure. This is due to the finite
flexibility of chains, as
calculation of the bond angle distribution reveals
that the chains are bent slightly more 
than in the initial configuration.

We have also calculated the pair correlation 
function within the precursor layer corresponding 
to late stages of spreading.
The structure of the precursor film 
in this case is disordered and liquid--like. 
At very late stages the neighboring chains 
tend to orient themselves in the same direction.
This is revealed by calculating the distribution of
orientations. 
The final stages of spreading correspond to 
diffusively thinning monolayer where the chains 
prefer to be relatively straight.

The density profiles of Fig. 12 are 
characterised by a rounded overall shape with 
a peak corresponding to the chains 
that are not yet in the precursor film. This is
consistent with the observations made from
the configurations of Fig. 11.
Fig. 13 shows $w(t)$
for three different system sizes. A noteworthy 
feature is that the crossover 
towards late--time diffusive behavior is not as clear as in 
the case of dimers. For the largest system size we 
find that 
$w(t) \sim t^{0.8 \pm 0.1}$ within the initial 
``almost linear'' regime, 
but our data for this system 
size does not extend far enough to capture the 
second ``diffusive'' regime. For the smallest 
system size $N=505$ we find that in the ``almost 
linear'' regime $w(t) \sim t^{0.8 \pm 0.1}$ and in 
the ``diffusive'' regime $w(t) \sim t^{0.5 \pm 0.1}$. 
>From these data we estimate that 
$D_{e} \approx 5.0 \times 10^{-6} \mbox{m}^2/\mbox{s}$ 
and $D_{\ell} \approx 
7.0 \times 10^{-7} \,\mbox{m}^2/\mbox{s}$ which give
$D_{e} / D_{\ell} \approx 7$.

For the present case, we also find a good data collapse
using the scaling form of Eq. (16), with
$x = 0.9$ and $y = 0.9$. Fig. 
14 shows the scaling function 
$\Phi(z) \sim const.$ for $z \ll 1$, and $\Phi(z) 
\sim z^{1/2-x}$ for $z \gg 1$ \cite{Nie94}. 

\vspace{0.5cm}

{\bf Completely flexible tetramers}

\vspace{0.5cm}

The results for completely flexible tetramers are
very similar to the results for rather stiff ones.
However, in this case the density
profiles in Fig. 15 appear to be somewhat flattened at 
late stages as compared to the rather stiff case.
This is evidently due to the greater flexibility of
the chains. 
The width of the precursor film for $N=505$ tetramers
gives $w(t) 
\sim t^{0.9 \pm 0.1}$ in the ``almost linear'' regime,
and $w(t) \sim t^{0.5 \pm 0.1}$ in the ``diffusive''
regime. We find
$D_{e} \approx 4.5 \times 10^{-6} \,\mbox{m}^2/\mbox{s}$ 
and $D_{\ell} \approx 
2.6 \times 10^{-7} \,\mbox{m}^2/\mbox{s}$ which
give $D_{e} / D_{\ell} \approx 20$.
Scaling of Eq. (16) is again obeyed, with
$x = 0.9$ and $y = 0.9$.

\vspace{0.5cm}

{\bf Completely flexible tetramers -- shifted case}

\vspace{0.5cm}

For the shifted tetramer case, 
the grafted end has the usual surface 
potential $V_{\mbox{I}}$ whereas the other end had potential 
$V_{\mbox{V}}$ for which the equilibrium distance from the 
surface extends about three bond lengths further away. 
We thus expect the results to be similar to those for
the shifted dimers, with possible differences arising from
the greater configurational entropy of the chainlike molecules.
Here we mostly concentrate in the morphology of the 
droplets during spreading.

Figs. 16(a) and (b) show the evolution of the droplet for $N=1010$. 
It can be seen that the evolution is 
strikingly different from the ordinary case.
As in the case of shifted dimers, the 
precursor films appears to be very compact and well--ordered. 
We have calculated the pair 
correlation function within the precursor film, which is 
shown in Fig. 17. Clear peaks corresponding up to about 
fourth or fifth nearest neighbour chains are clearly present, 
which is an indication of a high degree of local ordering. 
If one compares the pair correlation functions between 
shifted dimers and tetramers, it can be seen that for 
tetramers the peaks appear to be somewhat broadened. This 
is evidently due to the chain--like structure of the 
tetramers. Thus, with the present choice of interactions
the influence of chain flexibility is rather small even
at high temperatures.

We have also followed the time evolution of the density 
profile of the droplet. This is shown in Fig. 18. 
The profile develops from 
a fairly sharply peaked one towards a smooth but 
compact form. These profiles again confirm the observations 
made from the snapshots. 
We have estimated the effective diffusion 
barriers at late times
and find that $E_{a}' \approx 
2.9\,\mbox{eV}$ compared to $\approx 1.2 \,\mbox{eV}$ for the 
shifted dimer case. This can be understood on the basis that
within a well--ordered layer, the 
number of neighbors for tetramers should 
be roughly more than twice the corresponding
number for dimers.

\subsection{Octamers}

The results for chains built up of eight monomers (octamers) 
are presented in this section. We have studied two 
different systems, namely one in which the chains 
are rather stiff and another in which the chains are 
completely flexible. The grafted end has the usual surface 
potential $V_{\mbox{I}}$ whereas the other monomers 
have $V_{\mbox{IV}}$.
It should be pointed out that due to CPU--time constraints 
out system sizes are relatively small.

\vspace{0.5cm}

{\bf Rather stiff octamers}

\vspace{0.5cm}

Initial configuration for a droplet of rather stiff 
octamers ($\epsilon_{\theta}=10 \epsilon_f$) of size 
$N=488$ is shown in Fig. 19(a). It is characterised by a 
fairly complex structure. The tendency of the chains 
to be relatively straight is clearly visible. Fig. 
19(b) shows a typical evolution of the droplet. 
The structure in the middle of the precursor 
film is very complex. Chains 
at the edges of the droplet again
tend to lie parallel 
to the surface. 
We have calculated different time--dependent 
quantities such as the pair correlation function within 
the precursor film taken at 
late stages of spreading. From the shape of the function 
we can immediately conclude 
that the precursor film is disordered 
and liquid--like. 
The directions 
of neighbouring chains are fairly strongly correlated,
however. 
Calculation of the distribution of bond angles 
reveals that the 
chains tend to remain fairly straight through the whole 
spreading process. 

A set of typical
density profiles are shown in 
Fig. 20. It can be seen 
that at later times they become rather 
smooth and rounded. 
For the width of the precursor film 
we find that initially $w(t) \sim t^{0.9 \pm 0.1}$ with
$D_{e} \approx 1.7 \times 10^{-5} \,\mbox{m}^2/\mbox{s}$.
For the late--time behavior, we find
that $w(t) \sim t^{0.5 \pm 0.1}$ and 
$D_{\ell} \approx 4.3 \times 10^{-6} \,\mbox{m}^2/\mbox{s}$.
For the ratio we thus find $D_{e} / D_{\ell} \approx 4$.  
As far as the scaling is concerned, we were not able 
to collapse the data for different system sizes for
the present case. 

\vspace{0.5cm}

{\bf Completely flexible octamers}

\vspace{0.5cm}

The results for $N=488$ completely flexible octamers are
very similar to results for the rather stiff ones.
Again the precursor film is disordered and a 
crossover from ``almost linear'' ($w(t) \sim t^{0.8\pm 0.1}$)
towards ``diffusive'' ($w(t) \sim t^{0.5 \pm 0.1}$) behavior
for $w(t)$ is recovered, with 
$D_{e} \approx 3.7 \times 10^{-5} \,\mbox{m}^2/\mbox{s}$ and
$D_{\ell} \approx 1.77 \times 10^{-6} \,\mbox{m}^2/\mbox{s}$,
which yield $D_{e} / D_{\ell} \approx 20$.
The density profiles
shown in Fig. 21 bear a close resemblance to
the ones for rather stiff octamers, the late--time profiles
being somewhat less rounded in this case.

\subsection{The influence of Langevin dynamics}

We have also implemented Brownian dynamics into our 
computer code, the motivation being to study the
influence of a local thermostat for the present
case. Recently, the NH thermostat has been claimed
to be physically unsuitable for microscopic 
studies of droplet spreading \cite{DeC95}, despite
the fact that for a small system out
of equilibrium, there is no unique ``best'' choice.
We note that since we have a smooth surface, coupling
to an auxiliary thermostat must be used here.

To check our results,
we have performed additional 
simulations for $N=555$ shifted dimers \cite{Haa95II}.
We have employed two different values for 
the friction coefficient $\eta$, namely $\eta_1=3 \times 
10^{14} \,\mbox{s}^{-1}$ and $\eta_{2}=0.3 \times 
10^{14} \,\mbox{s}^{-1}$. We have set our bare time 
step to $0.01 \times 10^{-14}\,\mbox{s}$. For 
$\eta_1$ we recover the two different regimes 
for $w(t)$ with $w(t) \sim t^{1.0 \pm 0.1}$ and 
$w(t) \sim t^{0.5 \pm 0.1}$. We can again extract 
the associated ``diffusion'' coefficients with 
the result that $D_{e} \approx 5 \times 
10^{-6} \,\mbox{m}^2/\mbox{s}$ and $D_{\ell} \approx 1.5 
\times 10^{-7} \,\mbox{m}^2/\mbox{s}$ for the early and late 
time regime, respectively.
For the ratio we thus find $D_{e} / D_{\ell} \approx 30$.
Calculation of the pair correlation function 
within the precursor film again reveals that 
the layer has a high degree of local order. 

For $\eta_{2}$ we also recover the two regimes 
with $w(t) \sim t^{0.9 \pm 0.1}$ and $w(t) \sim t^
{0.4 \pm 0.1}$. For this particular case
$D_{e} \approx 8 \times 
10^{-5} \,\mbox{m}^2/\mbox{s}$ and $D_{\ell} \approx 4.0 
\times 10^{-7} \,\mbox{m}^2/\mbox{s}$ for the early and late 
time regimes, respectively.
For the ratio we obtain $D_{e} / D_{\ell} \approx 200$.
Reducing the friction would further increase this ratio 
since the late--time regime is dominated by
effective diffusion barriers that are independent of
$\eta$.

Based on our additional results we can conclude that
qualitatively {\it and}
quantitatively our results and conclusions
are unaffected by the choice of the thermostat. 
The main effect of the local thermostat with respect
to the NS thermostat is 
a slightly smoother crossover towards 
the late--time regime.

\section{Conclusions and discussion}

In this work we have studied a simple model of
dynamics of spreading for 
rigid and flexible molecules that interact
asymmetrically with respect to a solid surface.
We have studied two different cases. In the
ordinary case, the end potentials are of
different strength, but the equilibrium position
of the molecules on the surface 
is horizontal. In the shifted
case, however, the other end of the molecule
has an equilibrium distance that is compatible
with the length of the chain, {\it i.e.} the equilibrium
position is vertical with respect to the surface.
The latter case in particular
can be considered as an effective
model for the case of 
hydrophobic and hydrophilic surface 
interactions \cite{Caz94}.

One of our main results is that the microscopic
structure in the precursor film drastically depends
on the nature of the asymmetrical interactions.
For the shifted case, there is a high degree of
local order present which makes the density profile
of the droplet unusually sharp and flat. This result
is in good qualitative agreement with a recent experiment
on a physically similar system \cite{Caz94}.
Moreover, our model recovers the overall $t^{0.5}$
time dependence of the radius of the precursor
film, and we have been able to quantitatively
estimate the associated transport coefficients.
Typical ratios of the early--time coefficients $D_e$
to the late--time ones $D_{\ell}$ are in very good 
agreement with the experiments. Furthermore, our
model demonstrates how this ratio increases with
increasing local order in the precursor film,
in cases where the late--time diffusion is controlled
by energy barriers arising from neighboring molecules. 

Recently, the choice of the global NH thermostat used
here and in Refs. \cite{Nie92,Nie94} 
has been criticized by 
De Coninck {\it et al.} \cite{DeC95} on basis of the argument that 
in an inhomogeneous system a global thermostat is not physically 
justified.
However, due to our smooth surface the heat must dissipated
by other means than coupling to substrate atoms held at 
constant temperature, as was done
in Ref. \cite{DeC95}. 
Moreover, it is a well known fact that there is no unique way
of controlling the instantaneous temperature of
a non--equilibrium system. We have performed our
simulations mainly with the NH thermostat, 
but test runs with Brownian dynamics reveal
that the same qualitative and quantitative behavior
persists. Moreover, the results of Ref. \cite{Nie94}
as well as those of the present work 
compare very well with the
simulations of De Coninck {\it et al.} 
\cite{DeC95},
once differences in the geometries (cylindrical vs.
spherical) are properly
accounted for. De Coninck {\it et al.}
found that for largest
droplets the number of atoms in the first ({\it i.e.} the
precursor) layer was well 
described by $N(t) \sim t^{0.85 \pm 0.05}$
and the corresponding radius by 
$R^2(t) \sim t^{0.82 \pm 0.06}$ \cite{DeC95}.
On the other hand, in our geometry $w(t) \sim N(t) \sim
t^{0.9\pm 0.1}$ for the ``almost linear'' regime. 
The flux of particles into the precursor layer is the same
in the two independent studies and therefore the 
results are equivalent. The fact that the slower late--time
``diffusive'' regime was not reported in Ref. \cite{DeC95}
is probably due to insufficient simulation times, since
their system sizes were rather large.
Thus both
the qualitative and quantitative features of the
spreading phenomenon are fairly insensitive to the
choice of thermostat as well as the geometry.

To summarize, we hope to have further demonstrated in
this work that the spreading phenomenon at microscopic 
length--scales is a very complicated process. 
It seems highly unlikely that the properties of all the different
cases studied here and in other works could be obtained 
from a more general framework. On the other hand, there are
many features of spreading, such as the time--dependence
of the precursor radius that are rather insensitive to the
details of interactions, or molecular structure of the liquid.
This work calls
for more systematic and controlled experimental
as well as theoretical 
work in order to further classify the properties of 
tiny liquid droplets
spreading on a solid surface.

\clearpage

\begin{table}[t]
  \begin{center}
  \vspace{0.5cm}
  \begin{tabular}{|c|c|c|}
  \multicolumn{3}{|c|}{Surface interaction parameters}
     \\ \hline
  $\epsilon_{i}$      &   $\sigma_{i}$   &    Symbol
     \\ \hline
  1.0 $\epsilon_{f}$  & 5.0 $\sigma_{f}$ &    $V_{\mbox{I}}$
      \\ \hline
  0.06 $\epsilon_{f}$ & 5.0 $\sigma_{f}$ &    $V_{\mbox{II}}$
      \\ \hline
  0.02 $\epsilon_{f}$ & 7.3 $\sigma_{f}$ &    $V_{\mbox{III}}$
      \\ \hline
  0.006 $\epsilon_{f}$ & 5.0 $\sigma_{f}$ &   $V_{\mbox{IV}}$
      \\ \hline
  0.01 $\epsilon_{f}$ & 8.0 $\sigma_{f}$ &    $V_{\mbox{V}}$
      \\ 
  \end{tabular}
  \vspace{0.5cm}
  \caption{Surface interaction parameters and their symbols 
used in this study.}
  \vspace{0.5cm}
  \end{center}
\end{table}

\cleardoublepage
\pagebreak
\begin{center}
\Large 
{\sc figure captions}
\end{center}

\normalsize

\vspace{2.0cm}

Fig. 1. Schematic illustration of the different surface potentials
used in this study. The ``grafted'' end 
always has potential $V_{\rm I}$.

\vspace{1.0cm}

Fig. 2. (a) Initial configuration for the ordinary 
case of $N=1525$ dimers. 
The grafted end
is represented by a large filled circle.
(b) Same system taken at $ t = 30000\, \mbox{r.u.}$.

\vspace{1.0cm}

Fig. 3. Pair correlation function within the precursor film
at $ t = 80000\, \mbox{r.u.}$. Note the
disordered and liquid--like structure of the film.

\vspace{1.0cm}

Fig. 4. Density profiles for the case of ordinary
dimers. The late--time shoulders are
due to the dimers that lie flat on the surface.
These and all the other density profiles have been smoothed
to remove noise.
\vspace{1.0cm}

Fig. 5. The width of the precursor film $w(t)$ 
for the ordinary dimer case. It can
be characterised by an ``almost linear'' 
regime which crosses over to a ``diffusive'' one.

\vspace{1.0cm}
Fig. 6. (a) Initial configuration for the shifted 
case of $N=1525$ dimers. 
(b) Same system taken at $ t = 30000\, \mbox{r.u.}$. 

\vspace{1.0cm}
Fig. 7. Pair correlation function within the precursor film
at $ t = 80000\, \mbox{r.u.}$. Clear
peaks can be observed corresponding up to fourth or fifth nearest
neighbour, indicating that the film displays a high degree of local
order.

\vspace{1.0cm}
Fig. 8. Distribution of orientations $d_o$ 
for the shifted dimer case taken
at $t=80000 \,\mbox{r.u.}$. Orientations of 
dimers are correlated over
the whole system.

\vspace{1.0cm}
Fig. 9. (a) Density profiles for the shifted dimer case.
(b) Experimental density profiles for 
triloxane polyoxyethylene molecules 
(hydrophobic head with a hydrophilic tail)
spreading on silica bearing a dense grafted layer
of trimethyls \cite{Caz94}. The profiles are
strikingly similar.

\vspace{1.0cm}
Fig. 10. $w(t)$ for the shifted dimer case. It can
be characterised by an ``almost linear'' regime which crosses over to
``diffusive'' one. The late--time ``diffusion'' coefficient is
roughly one order of magnitude smaller than for the ordinary case.

\vspace{1.0cm}
Fig. 11. (a) The initial configuration for 
$N=785$ rather stiff tetramers. 
(b) Same system taken at $t=80000 \, \mbox{r.u.}$. Notice the complicated
and disordered structure of the precursor filmi due to the flexibility
of the chains.

\vspace{1.0cm}
Fig. 12. Density profiles for rather stiff tetramers taken at three
different times. The profiles are fairly smooth and rounded.

\vspace{1.0cm}
Fig. 13. $w(t)$ for rather stiff tetramers with three different
system sizes. 

\vspace{1.0cm}
Fig. 14. Scaled data for $w(t)$ for the three different system sizes
with $x=0.9$ and $y=0.9$.

\vspace{1.0cm}
Fig. 15. Density profiles for completely flexible 
tetramers taken at three different times. 
The profiles are somewhat flatter as compared
to the rather stiff case due to the flexibility of the
chains.

\vspace{1.0cm}
Fig. 16. (a) Initial configuration for the shifted case of $N=1010$ 
completely flexible tetramers. 
(b) Same system taken at $t=80000 \, \mbox{r.u.}$. 
Notice the appearance
of a compact precursor film with sharp edges.

\vspace{1.0cm}
Fig. 17. Pair correlation function within the precursor film
at $ t = 80000\, \mbox{r.u.}$. Clear
peaks can be observed corresponding up to fourth or fifth nearest
neighbour, indicating that the film displays a high degree of local
order. 

\vspace{1.0cm}
Fig. 18. Density profiles for the shifted case of completely
flexible tetramers taken at three
different times. 

\vspace{1.0cm}
Fig. 19. (a) Initial configuration for the case of $N=488$ 
rather stiff octamers. 
(b) Same system taken at $t=80000 \, \mbox{r.u.}$. The structure of the
precursor film is again disordered and liquid--like.

\vspace{1.0cm}
Fig. 20. Density profiles for the case of rather stiff 
octamers taken at three different times. 
The profiles are fairly smooth and rounded.

\vspace{1.0cm}
Fig. 21. Density profiles for the case of completely flexible 
octamers taken at three different times. 
The profiles are somewhat flatter than in the case
of rather stiff octamers.


\begin{thebibliography}{99}
\bibitem{diffusive}
This holds, provided that the ratio of the 
radius of the precursor
film and the radius of the second layer remains approximately
constant.
\end{thebibliography}
\end{document}